# Electric field induced strong enhancement of electroluminescence in multi-layer MoS$_2$

Dehui Li[1], Rui Cheng[2], Hailong Zhou[1], Chen Wang[2], Anxiang Yin[1], Yu Chen[2], Nathan O. Weiss[2], Yu Huang[2,3], Xiangfeng Duan[1,3,*]

[1]Department of Chemistry and Biochemistry, University of California, Los Angeles, CA 90095, USA; [2]Department of Materials Science and Engineering, University of California, Los Angeles, CA 90095, USA; [3]California Nanosystems Institute, University of California, Los Angeles, CA 90095, USA

*Corresponding email: xduan@chem.ucla.edu

**The layered transition metal dichalcogenides (TMDs) have attracted considerable interest due to their unique electronic and optical properties. Here we report electric field induced strong electroluminescence in multi-layer MoS$_2$ and WSe$_2$. We show that GaN-Al$_2$O$_3$-MoS$_2$ and GaN-Al$_2$O$_3$-MoS$_2$-Al$_2$O$_3$-graphene vertical heterojunctions can be created with excellent rectification behaviour. Electroluminescence studies demonstrate prominent direct bandgap excitonic emission in multi-layer MoS$_2$ over the entire vertical junction area. Importantly, the electroluminescence efficiency observed in multi-layer MoS$_2$ is comparable to or even higher than that in monolayers, corresponding to a relative electroluminescence enhancement factor of >1000 in multi-layer MoS$_2$ when compared to its photoluminescence. This striking enhancement of electroluminescence can be attributed to the high electric field induced carrier redistribution from low energy points (indirect bandgap) to high energy points**

**(direct bandgap) of *k*-space, arising from the unique band structure of MoS$_2$ with a much higher density of states at high energy points. The electric field induced electroluminescence is general for other TMDs including WSe$_2$, and can provide a fundamental platform to probe the carrier injection, population and recombination in multi-layer TMDs and open up a new pathway toward TMD based optoelectronic devices.**

The layered transition metal dichalcogenides (TMDs) have attracted considerable interest for their unique layer-number dependent electronic and optical properties[1-15]. The monolayer MoS$_2$ and WSe$_2$ exhibit a direct bandgap with strong photoluminescence and are of particular interest for new types of optoelectronic devices[2,9,10,12,16-20]. However, the multi-layer MoS$_2$ is an indirect bandgap semiconductor and generally believed to be optically inactive, with the photoluminescence typically more than 3 orders of magnitude weaker than that of the monolayers[2,7], which prevents the multi-layer MoS$_2$ from being used in the field of light emitting devices.

To achieve efficient electroluminescence (EL), a p-n junction is usually required to simultaneously inject electrons and holes that recombine to give out photon emission. In case of two-dimensional transition metal dichalcogenides (TMDs), it is possible to create either a lateral junction with a one dimensional (1D) junction interface or a vertically stacked structure with a two-dimensional (2D) junction interface for electron or hole injection that are necessary for EL emission. The EL emission from the lateral junctions of mono-/few-layer TMDs has been recently reported and is typically limited to the local area across the 1D junction interface[21-27] because the full depletion of the monolayers prevents efficient carrier injection throughout the entire monolayer. In order to achieve broad area EL emission, the vertically stacked structures with 2D junction interface are usually preferred[5]. However, with atomically thin thickness, the rapid carrier leakage[28] across the monolayer TMDs in the vertical junctions could prevent the efficient recombination of the injected carriers and limit the EL efficiency in the vertically stacked electrically driven light emitting devices. To this end, multi-layer TMDs may be beneficial, which however are typically indirect bandgap semiconductors and normally optically inactive. Here we report, for the first time, strong EL emission from the entire 2D junction interface in a unique design of vertical heterostructure devices with multi-layer $MoS_2$ and $WSe_2$, (up to ~130 layers), and demonstrate a strong electric field induced enhancement of the electroluminescence in indirect bandgap multi-layer $MoS_2$ and $WSe_2$ to achieve comparable or better efficiency than that of direct bandgap monolayer counterparts.

**Results**

**Fabrication and Structural characterization of $GaN$-$Al_2O_3$-$MoS_2$ structures**. To create the vertically stacked light emitting devices based on n-type $MoS_2$, a p-type doped GaN is used to inject holes into $MoS_2$ flakes (Fig. 1a, b). A thin layer of 4 nm thick $Al_2O_3$ was deposited on the p-type GaN (~3.8 μm p-GaN on sapphire) before transferring of the mechanically exfoliated $MoS_2$ flakes (see methods). The insulating $Al_2O_3$ layer can partly block the electrons to inject from n-$MoS_2$ to p-GaN while allowing holes to be effectively injected from p-GaN to n-$MoS_2$ due to the unique band alignment (see band diagram in Fig. 1f, g and further discussion below), where the desired EL occurs[29]. Additionally, the insulating layer can also suppress the direct tunneling of electrons or holes from the top electrode to GaN substrate (through the thin $MoS_2$ flakes) to improve the EL efficiency. A Ni/Au metal thin film was used as the top contact electrodes at the edge of the MoS$_2$ and the p-GaN substrate was contacted with a Pd/Au metal thin film (Fig. S1). Figure 1c displays an optical image of the top view of a typical device on the GaN substrate, where the MoS$_2$ flake and electrodes are indicated. A high resolution cross-sectional transmission electron microscope (TEM) image clearly shows the interfaces of the GaN-Al$_2$O$_3$-MoS$_2$ vertical stack (Fig. 1d).

**Band alignment and output characteristics of GaN-Al$_2$O$_3$-MoS$_2$ structures.** Figures 1 e, f, g show the ideal band diagrams of the vertical heterostructures. At or near the zero bias, the insulating layer blocks both electrons and holes from passing through the heterojunction with a zero current (Fig. 1e). It should be noted that the valence band energy difference between GaN and Al$_2$O$_3$ ($\Delta E_v$=1 eV) is much smaller than that of the conduction bands ($\Delta E_c$=2.2 eV). As a consequence, it is much easier for holes to be injected from GaN to MoS$_2$ flakes due to a relatively lower potential barrier. Applying a positive voltage to p-type GaN will drive the holes to tunnel from GaN to MoS$_2$, resulting in a forward current (Fig. 1e). Further increasing the applied positive voltage on the GaN, the holes in GaN can even thermally emit into MoS$_2$, with the injection current rapidly increasing with the bias voltage. Under sufficient high voltage, electrons in MoS$_2$ are able to tunnel to the conduction band of the GaN, which can also contribute to the forward current and lead to the emission from the GaN substrate. Under a negative bias, the top of valence band of the GaN falls within the bandgap of MoS$_2$ while the bottom of the conduction band of MoS$_2$ falls within the bandgap of GaN (Fig. 1g). Therefore, there is no available state for both electrons and holes to tunnel through the insulating layer, resulting in zero current. The current voltage output characteristic (Fig. 1h) shows clear current rectification behavior, demonstrating excellent diode characteristics in the GaN-Al$_2$O$_3$-MoS$_2$ vertical devices to satisfy the basic requirement for the high efficiency EL devices.

**Electroluminescence in GaN-Al$_2$O$_3$-MoS$_2$ structures.** Importantly, both the monolayer and multi-layer MoS$_2$ can exhibit clear EL emission under a forward bias. The EL spectra of a monolayer and a 50-layer MoS$_2$ device are shown in Figure 2a and 2b. To identify and assign the origin of the EL peaks, the PL spectra of GaN and MoS$_2$ are displayed as well. Close comparison of the EL spectra with the PL spectra, we can unambiguously assign the EL emission peak around 530 nm to the defect emission of the GaN substrate, the peak at 667 nm to the exciton A emission in the monolayer MoS$_2$, the peaks at 691 nm and 635 nm to the exciton A and B emission in the 50-layer MoS$_2$ device[30]. The small periodic oscillations can be attributed to the interference patterns formed due to the GaN substrate. For both monolayer and multi-layer MoS$_2$ devices, the EL peaks show a small redshift compared with the corresponding PL peaks due to self-heating effect, which is commonly seen in EL spectra of traditional semiconductor heterostructures[31]. It is also noted that the full width at half maximum (FWHM) of EL peaks of MoS$_2$ is much broader than that of corresponding PL peaks, which may be attributed to the electric field induced peak broadening[32]. To focus on the EL only from MoS$_2$, we insert a long pass filter with a cut off wavelength of 650 nm to eliminate the emission from GaN substrate. Figure 2c and d shows the optical image (left panel) and EL mapping (right panel) for the monolayer and the 50-layer MoS$_2$ flakes. The profile of MoS$_2$ flakes and the contact electrodes are outlined by dashed lines in order to identify the position of the EL. Importantly, bright EL are clearly seen from entire overlapping area of the vertical heterostructures for both the monolayer and multi-layer MoS$_2$ flakes. The non-uniform distribution of the EL may be attributed to the imperfect interfaces (or impurities) due to the wet transfer process, the surface roughness of p-GaN substrate and/or non-uniform field distribution.

Next we have investigated the thickness dependence to probe the influence of thickness on the EL efficiency and understand the mechanism for the enhanced EL in the multi-layer indirect bandgap $MoS_2$. To properly compare the EL efficiency among different devices, we have normalized the EL spectra by the injection current density, which is proportional to the EL efficiency. Interestingly, the overall normalized intensity for all devices with different $MoS_2$ thickness (including the monolayer and thicker one up to 50 layers) is highly comparable with each other (Fig. 2e and Supplementary Fig. S2). Both the exciton A and B can be identified in the EL spectra of $MoS_2$ flakes thicker than 6 nm, whereas only exciton A emission can be clearly seen from EL spectra of the thinner flakes, which may be ascribed to the stronger electric field induced peak broadening in the thin layers. In the thinner flakes, the electric field is much stronger and the field induced peak broadening is so severe that the exciton A and B emission merge into one broad peak[32].

The observation of nearly comparable EL intensity in monolayer and multi-layer $MoS_2$ is quite striking when considering the PL intensity is strongly dependent on the number of layers in these TMD materials. With a direct bandgap, monolayer can typically exhibit rather strong PL, which deceases rapidly with the increasing number of atomic layers due to the crossover from a direct bandgap semiconductor to an indirect one with increasing layer number[2,7]. For PL process, the photogenerated carriers undergo rapid thermalization and reach the thermal equilibrium in a time scale much shorter than that of radiative recombination[33], with the majority of carriers occupying their immediate band extremum, *i. e.* electrons at *Λ* valley and holes at *Γ* hill of *k*-space in multi-layer MoS$_2$. As a result, the direct bandgap recombination at *K* point is rather weak due to the lack of carrier occupation at *K* point in the multi-layer MoS$_2$ under the thermal equilibrium. As the thickness increases, less and less carriers occupy the *K* point due to the increasing energy difference between the *K* valley and *Λ* valley for electrons ($\varDelta E_{\varLambda\text{-}K}$) and between the *K* hill and *Γ* hill for holes ($\varDelta E_{\varGamma\text{-}K}$), resulting in an exponential decrease of the PL intensity with the increase of MoS$_2$ thickness[7].

**Electric field induced carrier redistribution.** This distinct contrast between EL and PL is particularly evident in the plot of EL and PL *vs.* layer thickness, in which both EL and PL intensity observed in different layered samples are normalized by the EL and PL intensity of the monolayer MoS$_2$ (Fig. 2e). If we define the ratio of the EL intensity to PL intensity as a *relative electric field-induced EL enhancement factor*, the enhancement factor increases with the increasing MoS$_2$ layer number and reaches as large as ~2000 for a 50-layer sample (right axis of the Fig. 2f). It should be noted that the EL in all multi-layer MoS$_2$ originates from the direct bandgap (*K-K*) transition rather than the near band edge (*Λ-Γ*) transition. This can occur due to the fact that the electric field could induce carrier redistribution from low energy points to high energy points of *k*-space (Fig. 3a), leading to a non-equilibrium distribution of the injected carriers in the multi-layer indirect bandgap MoS$_2$. Under the applied electric field, the injected hot electrons and holes though the tunneling barriers are further accelerated and their kinetic energy is increased. Thus, the electron and hole temperature are significantly higher than that of the environment, causing electrons to transfer from *Λ* valley to *K* valley while holes to transfer from *Γ* hill to *K* hill[24,34]. In this way, electrons and holes can recombine at *K* point, leading to greatly enhanced direct bandgap EL in multi-layer MoS$_2$. Alternatively, the injected energetic hot carriers (through the tunneling barriers) can probably temporarily occupy the *K* valley and hill, which may contribute to the direct band gap EL as well. Nevertheless, the injected hot carriers can not normally sustain the high energy states (*K* valley and *K* hill) and will quickly decay to the lowest energy points (*Λ* valley for electrons and *Γ* hill for holes) before rediative recombination, because the hot carrier relaxation or cooling time in MoS$_2$ is on the order of sub-picosecond while the radiative recombination time is on the order of nanosecond[35,36]. Therefore, the enhanced EL should be primarily attributed to the electric-field induced redistribution of the injected carriers rather than the energetic hot carriers through the tunneling barriers.

It should be noted that the indirect bandgap exciton binding energy is rather large for bi-layer and tri-layer MoS$_2$ (around 400 meV for bi-layer and around 100 meV for tri-layer)[37], which can greatly suppress the electric field induced carrier redistribution. Consequently, EL in bi-layer and tri-layer MoS$_2$ is rather weak. When the thickness is larger than three layers, the indirect band gap exciton binding energy decreases to about 25 meV[37]. As a result, the indirect band gap excitons are mostly ionized at room temperature and can be efficiently redistributed to direct bandgap *K* points under high electrical field for EL emission. We have calculated the *K* point population fraction for electrons $n_2/(n_1+n_2)$ and holes $p_2/(p_1+p_2)$ as a function of the applied electric field for MoS$_2$ thicker than 3 L (Fig. 3b and c, and Supplementary Section 8). The *K* point population fraction for both electrons and holes starts to increase around 40 kV/cm and saturates at 1 MV/cm with a saturation value of ~0.8. Such a large fraction of population at higher energy direct bandgap *K* point is quite striking and largely responsible for the enhanced EL emission in multi-layer MoS$_2$, which is usually difficult to achieve in other indirect bandgap semiconductors (e.g. Si). This unique electric field enhanced EL can be attributed to the unique band structure of $MoS_2$: the density-of-states at $K$ point is much larger than that at $\Lambda(\Gamma)$ points due to the larger effective mass at $K$ valley(hill) and the larger number of equivalent hill numbers at $K$ points (6) compared with that at $\Gamma$ points (1)[38]. Indeed, the density-of-state ratio is evaluated to be 6.4 for holes between the $K$ hill and $\Gamma$ hill and 10.6 for electrons between $K$ valley and $\Lambda$ valley (Supplementary Section 8).

**Output characteristics of $GaN-Al_2O_3-MoS_2-Al_2O_3$-graphene structures.** With the $GaN-Al_2O_3-MoS_2$ structures, a metal electrode is deposited on the edge of $MoS_2$ flake for electron injection. Due to the resistive and depletion of $MoS_2$ along both the lateral and vertical direction, there are both vertical and lateral components of electric field. It is therefore difficult to precisely determine the exact electrical field and quantitatively correlate the experimental results with the theoretical ones. To further verify the proposed mechanism, we have created vertically stacked $GaN-Al_2O_3-MoS_2-Al_2O_3$-graphene heterostructures (Fig. 4a), in which graphene is used as the top electrode. In this case, the applied electric field is largely dominated by the vertical component. We have also included a 4-nm thick $Al_2O_3$ layer between the $MoS_2$ and graphene in the new heterostructures to prevent electrons and holes from directly transferring to graphene, which would quench the EL in $MoS_2$. The current voltage characteristic of this new structure also shows excellent diode behaviour (Fig. 4b). However, the injection current is typically several times smaller than that of previous $GaN-Al_2O_3-MoS_2$ structures at the same bias voltage due to the additional tunneling layer between $MoS_2$ and graphene.

**Electroluminescence in $GaN-Al_2O_3-MoS_2-Al_2O_3$-graphene structures.** The left panel of Figure 4c displays an optical image of a typical device with the $MoS_2$ flake composed of two parts with different thicknesses (lower part with a thickness of 36 nm and upper part with a thickness of 92 nm). The EL mapping (the middle panel of Fig. 4c) shows several important features: (i) EL is observed from the entire heterojunction area, confirming the formation of broad area vertical junction; (ii) the thicker part of the flake exhibits a stronger EL signal than the thinner area, which is in striking contrast to the PL mapping of the same sample (right panel of the Fig. 4c) in which the thinner part clearly shows a stronger PL emission. EL spectra further confirm that the EL intensity of the thicker part is around 1.5 times stronger than that of the thinner part under the same injection condition (Fig. 4d), while the corresponding PL intensity in the thicker part is <40% of that in the thinner area (Fig. 4e).

We have also investigated the EL characteristics as a function of injection current or vertical electric field. The EL intensity increases monotonously with the injection current and saturates around the injection current of 174 µA (Fig. 4f, g and Supplementary Fig. S3). It is also noted that the EL emission from GaN continues to increase with the increasing injection current after the $MoS_2$ emission reaches saturation (Supplementary Fig. S3), suggesting the EL saturation in $MoS_2$ under high injection current (field) is a unique feature of $MoS_2$. Overall, the EL efficiency in $MoS_2$ increases rapidly first and then decreases with the increasing injection current (Fig. 4g) or the increasing electric field (Fig. 4h). The vertical electric field is determined by the current density and carrier mobility ((Supplementary Section S6) using a space charge limited current model (Supplementary Section S7)

The trend of EL efficiency with increasing electric field may be attributed to two competing factors under high electric field that may affect the emission efficiency in the opposite way: the increase of the electron and hole population fraction at direct bandgap *K* point and increase of carrier temperature. To account for this trend, we calculated the relative EL efficiency under various electric field by taking into account both the electric field induced carrier redistribution and carrier temperature[39] (blue curve in Fig. 4h and Supplementary Section 8). In general, the electron, hole temperature and their population in the higher energy $K$ points all increase with the increasing electric field. The increase of the electron and hole population faction at direct bandgap $K$ point would result in a higher EL efficiency, while the increasing average carrier temperature could reduce the EL efficiency. Together, these two competing factors could lead to an increase of EL efficiency with electric field at lower field regime (where the rapid increase of carrier population at K point dominates the EL efficiency, see Fig. 3b, c and 4h), followed by an decrease of EL efficiency at higher electric field (where the carrier population at K point saturates, and the effect of the carrier temperature dominates the EL efficiency, see Fig. 3b, c and 4h). Importantly, the observed EL efficiency first increases and then decreases with the increasing electric field (red dots in Fig. 4h), which is consistent with the trend predicted by the theoretical calculations (blue line in Fig. 4h), except at low field (small injection current) regime where the space charge limited current model used to calculate the electric field underestimates the field in $MoS_2$ (Supplementary Section S7). The decrease in EL efficiency may also be partly attributed to the increased electron injection from $MoS_2$ to GaN substrate under high electric field[28], which is supported by the continued increase of GaN emission after the saturation of emission in $MoS_2$ under high electric field (Supplementary Fig. S3). Additionally, other factors such as field induced exciton ionization and exciton-exciton annihilation may also contribute to the decrease in EL efficiency under high injection current (field).

**Thickness dependence of the electroluminescence efficiency in GaN-Al$_2$O$_3$-MoS$_2$-Al$_2$O$_3$-graphene structures.** We have further probed the thickness dependence of the EL efficiency based on GaN-Al$_2$O$_3$-MoS$_2$-Al$_2$O$_3$-graphene heterostructures. The EL intensity normalized by the injection current density indicates that both exciton A and B peaks are present for all flakes with various thicknesses (6-92 nm) (Fig. 5a and Supplementary Fig. S4). The EL efficiency for the current structures is about 3 times higher than that in the previous GaN-Al$_2$O$_3$-MoS$_2$ structures, likely due to the carrier leakage through the metal electrode in previous structures (Supplementary Fig. S5). Importantly, a clear trend is observed that the normalized integrated EL intensity increases with the increasing flake thickness (Fig. 5b), which might be attributed to at least two factors. First, the applied electric field would be smaller in the thicker flakes for a fixed applied bias voltage, leading to a larger EL efficiency under the same injection current (see Fig. 4g, in the high field regime where the EL efficiency decreases with increasing electric field). The second factor is the injected carrier leakage[28]. The thicker the flake is, the smaller the percentage of the injected carrier can leak away from the active emitters, resulting in a stronger EL intensity. The theoretical calculation based on the above discussion agrees well with the experimental results (Fig. 5b and Supplementary Section 9). The relatively large discrepancy for the thinner flakes may arise from the inaccurate energy difference $\it{\Delta E_{\Lambda-K}}$ and $\it{\Delta E_{\Gamma-K}}$ and effective mass that can be different from the bulk values used in the calculation[40].

**Discussion**

After calibrating the collection efficiency of our EL measurement system, we have estimated the EL external quantum efficiency of our vertical heterostructure devices to be around 10$^{-4}$, which is about one order of magnitude larger than the EL efficiency in monolayer MoS$_2$ transistors[26] and on the same order as the reported value in monolayer WSe$_2$ planar structures[23]. We believed this enhanced EL in the multi-layer indirect bandgap MoS$_2$ flakes can be attributed to the unique energy band structure of MoS$_2$ and the vertical heterostructure design used in the current devices. The larger effective mass and the larger number of equivalent valleys and hills at *K* points give rise to a larger density-of-state at *K* point compared with that at *Λ* point and *Γ* point of *k*-space (Supplementary Section 8). Furthermore, the energy differences among these different valleys or hills ($\Delta E_{\Lambda\text{-}K}$ and $\Delta E_{\Gamma\text{-}K}$) are relatively small. The design of vertical heterostructure junction can create a strong vertical electric field inside the MoS$_2$ flakes, which induces efficient carrier redistribution from low energy points (*Λ*, *Γ*) to the high energy points (*K*) with a larger density-of-state, resulting in greatly enhanced EL in the multi-layer MoS$_2$ flakes. The EL efficiency may be further improved by using other TMDs with smaller energy difference $\Delta E_{\Lambda\text{-}K}$ and $\Delta E_{\Gamma\text{-}K}$, less bulk trapping states[23] or optimizing the device fabrication process to reduce interface impurities, defects and traps.

In summary, our studies for the first time report the *broad area* EL emission from the entire junction area of the vertically stacked heterostructures of MoS$_2$, and demonstrate *unusually strong* EL emission in the *indirect bandgap* multi-layer MoS$_2$. This unique EL characteristics can not be easily achieved in other traditional indirect semiconductors (e.g. Si), and is fundamentally originated from the unique electronic band structures of multi-layer MoS$_2$, and general for other TMD materials. Indeed, our preliminary studies indicate the electric field enhanced EL can also be observed in other multi-layer TMDs (e.g., WSe$_2$, see Supplementary Section 10 and Fig. S9). Our studies can thus not only offer a fundamental platform to probe the carrier injection, population and recombination in multi-layer TMDs under high electric field, but also open up a pathway toward new types of light emitting devices with spin- and valley-polarization[25] based on multi-layer TMDs.

**Methods**

**Device fabrication**

$MoS_2$ samples are mechanically exfoliated from a bulk $MoS_2$ crystal onto 285 nm $Si/SiO_2$ substrate with alignment markers and then transferred onto the p-GaN substrate with a doping level of $2\times10^{17}$ cm$^{-3}$ grown by metalorganic chemical vapour deposition (MOCVD). A layer of 4 nm thick $Al_2O_3$ was deposited on GaN substrate using atomic layer deposition (ALD) before $MoS_2$ flakes were transferred. Then a layer of 60 nm thick $Al_2O_3$ was deposited to insulate the substrate and top electrodes. Windows were defined by electron beam lithography followed by 6:1 buffered oxide etch to remove the $Al_2O_3$ on the top of $MoS_2$ samples and on GaN for depositing the contact electrodes. Finally, an electron beam evaporation process was employed to deposit 5-nm Ni/50-nm Au metal thin films as top contact electrodes at the edge of the $MoS_2$ and 5-nm Pd/50-nm Au as the bottom electrode on the GaN substrate to form Ohmic contact (Fig. S1). For vertically stacked GaN-$Al_2O_3$-$MoS_2$-$Al_2O_3$-graphene devices, a 4 nm thick $Al_2O_3$ was deposited after opening the windows. The graphene grown by CVD method was transferred onto the top of the windows. Then the transferred graphene was patterned by EBL followed by oxygen plasma etching so that the graphene only covers the $MoS_2$ flakes. Finally, 20-nm Ti/50-nm Au electrodes were defined on the edge of the graphene and 5-nm Pd/ 50-nm Au electrodes were deposited on the GaN substrate.

**Microscopic, electrical and optical characterizations**

The thickness of the MoS2 flakes was determined by tapping-mode atomic force microscopy (Vecco 5000 system). The cross-section image of the vertical devices was acquired in an FEI Titan high resolution transmission microscopy. The electrical measurement was carried out in a probe station (Lakeshore, TTP4) coupled with a precision source/measurement unit (Agilent B2902A). The PL measurement was conducted under a confocal mirco-Raman system (Horiba LABHR) equipped with a 600 gr/mm grating in a backscattering configuration excited by an Ar ion laser (514 nm) with a excitation power of 500 μW. The PL measurement of GaN substrate was excited by a 257 nm laser with a power of 200 μW.

**Electroluminescence measurement**

The EL measurement was carried out under a home-built micro-PL system (Acton 2300i spectrometer equipped with a 150 gr/mm grating) combining with a precision source/measurement unit (Agilent B2902A) to force voltage or current. The EL signal was collected using a 50X objective (N.A.=0.5) and acquired by liquid nitrogen cooled CCD (Princeton instruments PyLoN 400F).

**References**

1   Novoselov, K. S. *et al.* Two-dimensional atomic crystals. *Proc. Natl Acad. Sci. USA* **102**, 10451-10453 (2005).

2   Mak, K. F., Lee, C., Hone, J., Shan, J. & Heinz, T. F. Atomically thin MoS$_2$: A new direct-gap semiconductor. *Phys. Rev. Lett.* **105**, 136805 (2010).

3   Geim, A. & Grigorieva, I. Van der Waals heterostructures. *Nature* **499**, 419-425 (2013).

4   Yu, W. J. *et al.* Highly efficient gate-tunable photocurrent generation in vertical heterostructures of layered materials. *Nature Nanotech.* **8**, 952-958 (2013).

5   Lopez-Sanchez, O. *et al.* Light generation and harvesting in a van der Waals heterostructure. *ACS Nano* **8**, 3042-3048 (2014).

6   Cao, T. *et al.* Valley-selective circular dichroism of monolayer molybdenum disulphide. *Nature Commun.* **3**, 887 (2012).

7   Tongay, S. *et al.* Monolayer behaviour in bulk $ReS_2$ due to electronic and vibrational decoupling. *Nature Commun.* **5**, 3252 (2014).

8   Yu, W. J. *et al.* Vertically stacked multi-heterostructures of layered materials for logic transistors and complementary inverters. *Nature Mater.* **12**, 246-252 (2013).

9   Jones, A. M. *et al.* Spin-layer locking effects in optical orientation of exciton spin in bilayer $WSe_2$. *Nature Phys.* **10**, 130-134 (2014).

10  Wu, S. *et al.* Electrical tuning of valley magnetic moment through symmetry control in bilayer $MoS_2$. *Nature Phys.* **9**, 149-153 (2013).

11  Chhowalla, M. *et al.* The chemistry of two-dimensional layered transition metal dichalcogenide nanosheets. *Nat. Chem.* **5**, 263-275 (2013).

12  Wang, Q. H., Kalantar-Zadeh, K., Kis, A., Coleman, J. N. & Strano, M. S. Electronics and optoelectronics of two-dimensional transition metal dichalcogenides. *Nature Nanotech.* **7**, 699-712 (2012).

13  Radisavljevic, B., Radenovic, A., Brivio, J., Giacometti, V. & Kis, A. Single-layer $MoS_2$ transistors. *Nature Nanotech.* **6**, 147-150 (2011).

14  Lopez-Sanchez, O., Lembke, D., Kayci, M., Radenovic, A. & Kis, A. Ultrasensitive photodetectors based on monolayer $MoS_2$. *Nature Nanotech.* **8**, 497-501 (2013).

15  Roy, K. *et al.* Graphene-$MoS_2$ hybrid structures for multifunctional photoresponsive memory devices. *Nature Nanotech.* **8**, 826-830 (2013).

16  Xiao, D., Liu, G., Feng, W., Xu, X. & Yao, W. Coupled Spin and Valley Physics in Monolayers of $MoS_2$ and Other Group-VI Dichalcogenides. *Phys. Rev. Lett.* **108**, 196802 (2012).

17  Mak, K. F. *et al.* Tightly bound trions in monolayer $MoS_2$. *Nature Mater.* **12**, 207-211 (2013).

18  Britnell, L. *et al.* Strong light-matter interactions in heterostructures of atomically thin films. *Science* **340**, 1311-1314 (2013).

19  Britnell, L. *et al.* Field-effect tunneling transistor based on vertical graphene heterostructures. *Science* **335**, 947-950 (2012).

20    Withers, F. *et al.* Light-emitting diodes by band-structure engineering in van der Waals heterostructures. *Nature Mater.* **14**, 301-306 (2015).

21    Baugher, B. W. H., Churchill, H. O. H., Yang, Y. & Jarillo-Herrero, P. Optoelectronic devices based on electrically tunable p-n diodes in a monolayer dichalcogenide. *Nature Nanotech.* **9**, 262-267 (2014).

22    Pospischil, A., Furchi, M. M. & Mueller, T. Solar-energy conversion and light emission in an atomic monolayer p-n diode. *Nature Nanotech.* **9**, 257–261 (2014).

23    Ross, J. S. *et al.* Electrically tunable excitonic light-emitting diodes based on monolayer $WSe_2$ p-n junctions. *Nature Nanotech.* **9**, 268-272 (2014).

24    Jo, S., Ubrig, N., Berger, H., Kuzmenko, A. B. & Morpurgo, A. F. Mono- and Bilayer WS2 Light-Emitting Transistors. *Nano Lett.* **14**, 2019-2025 (2014).

25    Zhang, Y., Oka, T., Suzuki, R., Ye, J. & Iwasa, Y. Electrically Switchable Chiral Light-Emitting Transistor. *Science* **344**, 725-728 (2014).

26    Sundaram, R. S. *et al.* Electroluminescence in single layer $MoS_2$. *Nano Lett.* **13**, 1416-1421 (2013).

27    Cheng, R. *et al.* Electroluminescence and Photocurrent Generation from Atomically Sharp $WSe_2$/$MoS_2$ Heterojunction p–n Diodes. *Nano Lett.* **14**, 5590-5597 (2014).

28    Yamakoshi, S., Sanada, T., Wada, O., Umebu, I. & Sakurai, T. Direct observation of electron leakage in InGaAsP/InP double heterostructure. *Appl. Phys. Lett.* **40**, 144-146 (1982).

29    Min, K. W. *et al.* White-light emitting diode array of p+-Si/aligned n-$SnO_2$ nanowires heterojunctions. *Adv. Funct. Mater.* **21**, 119-124 (2011).

30    Splendiani, A. *et al.* Emerging photoluminescence in Monolayer $MoS_2$. *Nano Lett.* **10**, 1271-1275 (2010).

31    Fukai, Y. K., Matsuoka, Y. & Furuta, T. Measuring the junction temperature of AlGaAs/GaAs heterojunction bipolar transistors using electroluminescence. *Appl. Phys. Lett.* **63**, 340-342 (1993).

32    Li, D., Zhang, J., Zhang, Q. & Xiong, Q. Electric-Field-Dependent Photoconductivity in CdS Nanowires and Nanobelts: Exciton Ionization, Franz–Keldysh, and Stark Effects. *Nano Lett.* **12**, 2993-2999 (2012).

33	Pankove, J. I. *Optical processes in semiconductors*.  (Courier Dover Publications, 2012).

34	Sze, S. M. & Ng, K. K. *Physics of semiconductor devices*.  (John Wiley & Sons, 2006).

35	Nie, Z. *et al.* Ultrafast Carrier Thermalization and Cooling Dynamics in Few-Layer MoS2. *ACS Nano* **8**, 10931-10940 (2014).

36	Shi, H. *et al.* Exciton dynamics in suspended monolayer and few-layer MoS2 2D crystals. *ACS Nano* **7**, 1072-1080 (2013).

37	Komsa, H.-P. & Krasheninnikov, A. V. Effects of confinement and environment on the electronic structure and exciton binding energy of $MoS_2$ from first principles. *Phys. Rev. B* **86**, 241201 (2012).

38	Peelaers, H. & Van de Walle, C. G. Effects of strain on band structure and effective masses in $MoS_2$. *Phys. Rev. B* **86**, 241401 (2012).

39	Basu, P. K. *Theory of Optical Processes in Semiconductors: Bulk and Microstructures: Bulk and Microstructures*.  (Oxford University Press, 1997).

40	Yun, W. S., Han, S., Hong, S. C., Kim, I. G. & Lee, J. Thickness and strain effects on electronic structures of transition metal dichalcogenides: 2H-$MX_2$ semiconductors (M= Mo, W; X= S, Se, Te). *Phys. Rev. B* **85**, 033305 (2012).

**Figures & Legends**

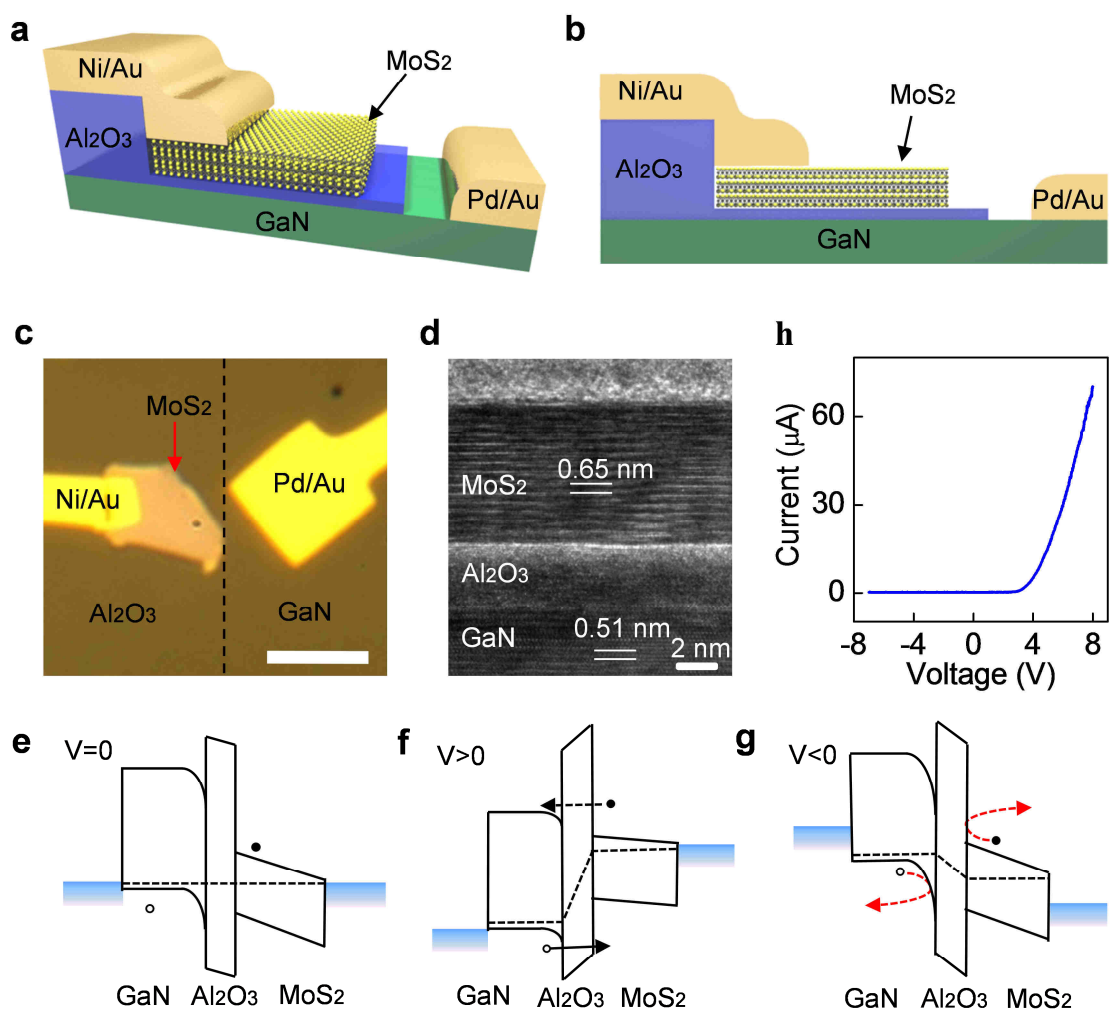

**Figure 1| Schematic illustration, structural and electrical characteristics, and band diagrams of the GaN-Al$_2$O$_3$-MoS$_2$ vertical devices. a,** A schematic of the three-dimensional view of the vertically stacked device. **b,** A schematic of the cross-sectional view of the device. **c,** An optical image of a GaN-Al$_2$O$_3$-MoS$_2$ vertical device. The dashed line highlights the area with Al$_2$O$_3$ layer and bare GaN surface. The scale bar is 4 μm. **d,** A cross-sectional high resolution TEM image of the interfaces across the GaN substrate, Al$_2$O$_3$ and MoS$_2$ flake vertical stack. The layer number of the MoS$_2$ flakes is 14. **e,** The ideal band diagram of the vertical heterostructure at zero bias. The dashed lines indicate the position of Fermi levels. At zero bias, the bottom of the conduction band and the top of valence band of MoS$_2$ fall within the forbidding bandgap of GaN. **f,** The ideal band diagram of the vertical heterostructure under forward bias. **g.** The ideal band diagram of the heterostructure under reverse bias. **h,** Current versus bias voltage characteristic of a vertically stacked device.

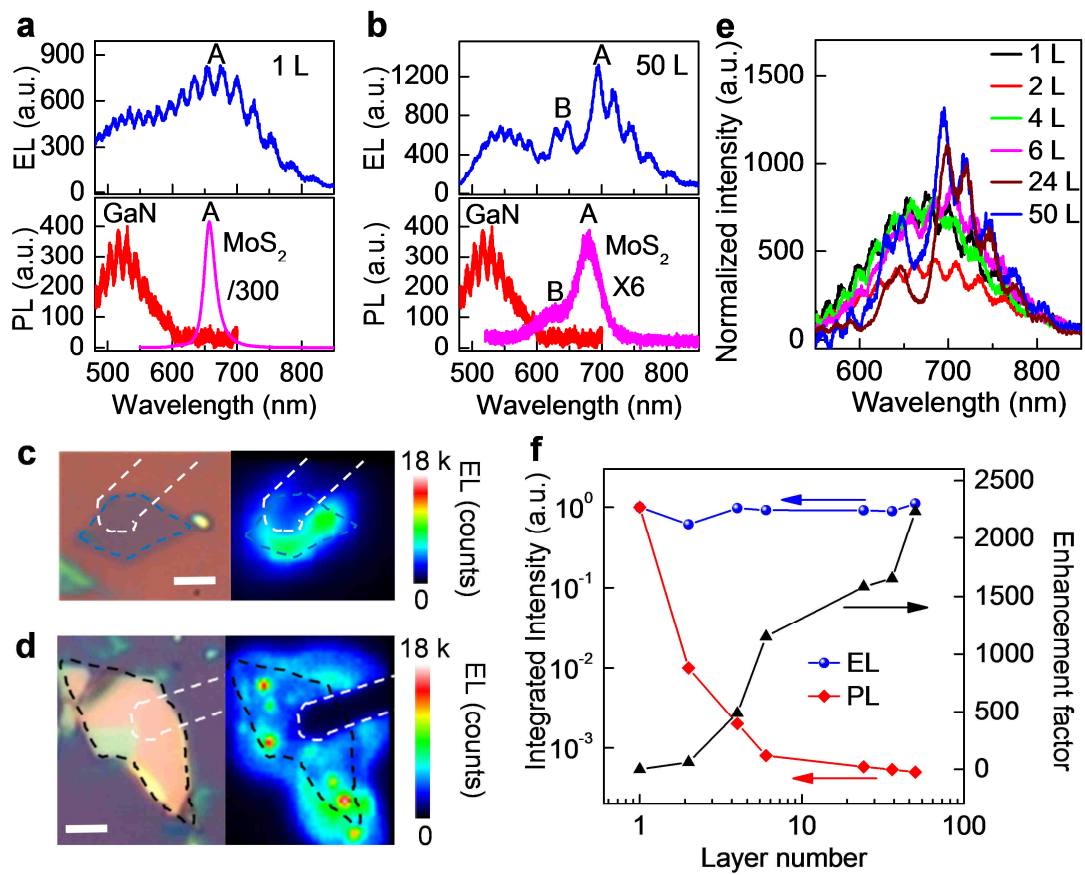

**Figure 2| Electroluminescence (EL) from the GaN-Al$_2$O$_3$-MoS$_2$ vertical devices. a,** The EL spectrum of a monolayer device under an injection current of 30 µA (upper panel). The PL spectra of GaN substrate and the same monolayer MoS$_2$ flake (divided by 300) are given as well in order to assign the EL peaks (bottom panel). **b,** The EL spectrum of a 50-layer MoS$_2$ device under an injection current of 88 µA (upper panel). The PL spectra of GaN substrate and the same 50-layer MoS$_2$ flake (multiplied by 6) are shown as well (bottom panel). The PL intensity of monolayer MoS$_2$ is around 2000 times larger than that of 50-layer MoS$_2$. **c,** The optical image (left panel) and EL mapping for the same monolayer device. The monolayer MoS$_2$ flake and electrode are outlined by dashed lines. A 650 nm long pass filter was used for mapping the emission from MoS$_2$ only. Scale bar: 3 µm. **d,** The optical image (left panel) and EL mapping for the same 50-layer MoS$_2$ device. The 50-layer MoS$_2$ flake and electrode are outlined by dashed lines. Scale bar: 3 µm. A 650 nm long pass filter was used for mapping the emission from MoS$_2$ only. **e,** The EL spectra from MoS$_2$ flakes with various number of layers. The spectra have been normalized by the injection current density in order to compare with each other and the GaN emission has been subtracted based on Gaussian fitting. **f,** The normalized integrated EL and PL intensity (left axis) and the relative enhancement factor defined as the ratio of EL intensity to the PL intensity (right axis) as a function of the layer number.

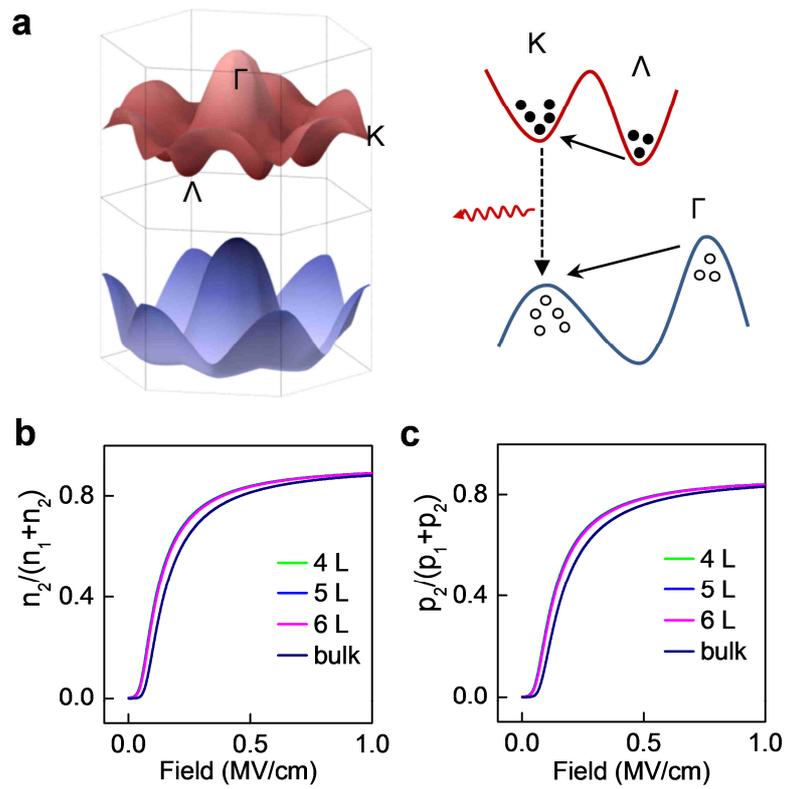

**Figure 3| The schematic of the carrier transfer processes and the calculated valley/hill population fraction. a,** The schematic illustration of the conduction band and valence band (left panel) and electric field induced carrier transfer between different energy valleys and hills (right panel). The equivalent valley number and hill number are 6 at *K* and *Λ* points and 1 at *Γ* point. **b,** The calculated electron population fraction at *K* valley as a function of the applied electric field for different thickness MoS$_2$ flakes. **c,** The calculated hole population fraction at *K* hill as a function of the applied electric field for different thickness MoS$_2$ flakes.

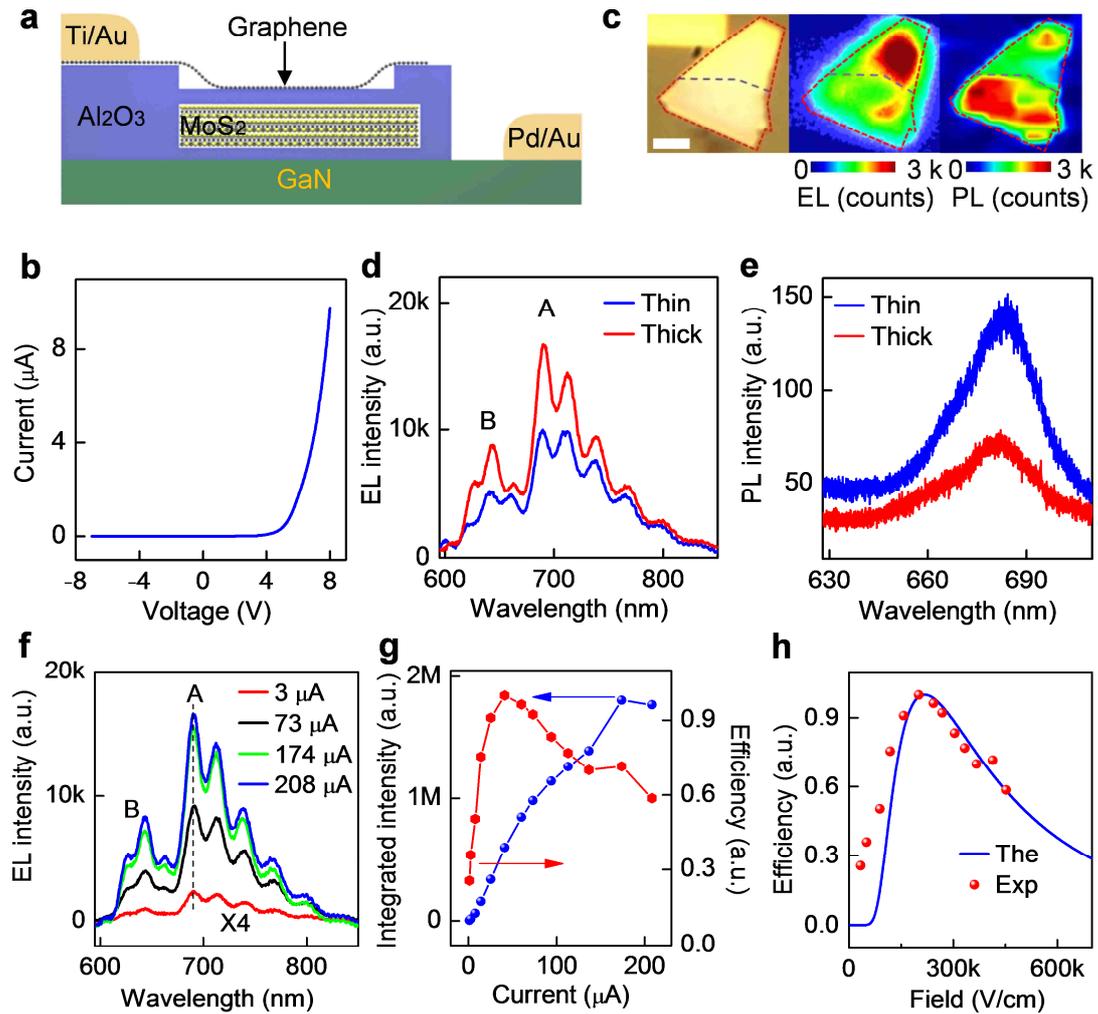

**Figure 4| EL from the vertically stacked GaN-Al$_2$O$_3$-MoS$_2$-Al$_2$O$_3$-graphene heterostructures. a,** A schematic illustration of the cross-sectional view of GaN-Al$_2$O$_3$-MoS$_2$-Al$_2$O$_3$-graphene vertical heterostructure. **b,** Current versus bias voltage characteristic of the GaN-Al$_2$O$_3$-MoS$_2$-Al$_2$O$_3$-graphene vertical device. **c,** The optical image (left panel), EL mapping (middle panel, under an injection current of 8 μA) and PL mapping (right panel, excited by a 514 nm Ar ion laser) of a multi-layer device. The MoS$_2$ flake is composed of two parts with different thicknesses: the 36 nm thick lower part and the 92 nm thick upper part. Scale bar: 3 μm. **d,** EL spectra from the thick part and thin part of the MoS$_2$ flake under an injection current of 174 μA. The GaN emission has been subtracted. **e,** PL spectra from the thick part and thin part of the same MoS$_2$ flake. **f,** EL spectra of the thick part at different injection current. The GaN emission has been subtracted. The corresponding applied voltage are 6 V, 13 V 17 V and 18 V. **g,** The integrated EL intensity and EL efficiency as a function of the injection current for the thick part. **h**, The EL efficiency as a function of the electric field. The discrete points are experimental results and the solid line is the theoretical calculation.

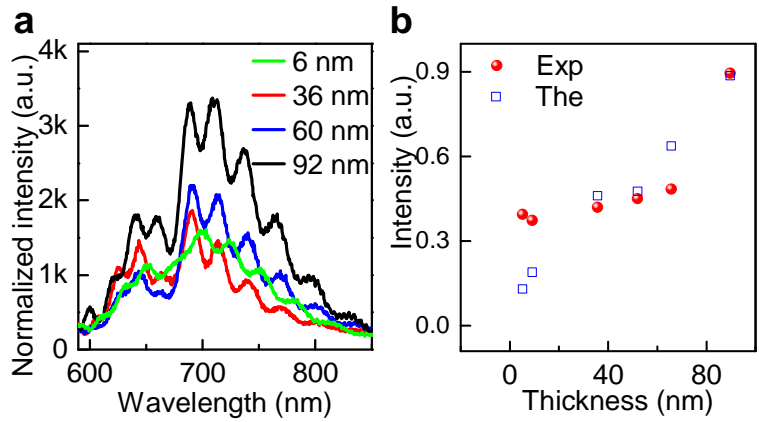

**Figure 5| Thickness dependent EL for the vertically stacked GaN-Al$_2$O$_3$-MoS$_2$-Al$_2$O$_3$-graphene heterostructures. a,** The normalized EL spectra for vertical devices with various thickness MoS$_2$ flake. The spectra are normalized by the injection current density in order to compare with each other. The GaN emission has been subtracted. **b,** The integrated EL intensity as a function of the thickness of the MoS$_2$ flakes. The red points are experimental results and the blue hollow squares are obtained from theoretical calculation.